\documentclass[twocolumn,showpacs,preprintnumbers,superscriptaddress,prl,amsmath,amssymb]{revtex4}

\usepackage{graphicx}
\usepackage{dcolumn}
\usepackage{bm}

\begin{document}

\title{Fragile structural transition in Mo$_3$Sb$_7$}

\author{J.-Q. Yan}
\affiliation{ Materials Science and Technology Division, Oak Ridge National Laboratory, Oak Ridge, TN 37831}
\affiliation{Department of Materials Science and Engineering, University of Tennessee, Knoxville, TN 37996}

\author{M. A. McGuire}
\affiliation{ Materials Science and Technology Division, Oak Ridge National Laboratory, Oak Ridge, TN 37831}

\author{A. F. May}
\affiliation{ Materials Science and Technology Division, Oak Ridge National Laboratory, Oak Ridge, TN 37831}

\author{D. Parker}
\affiliation{ Materials Science and Technology Division, Oak Ridge National Laboratory, Oak Ridge, TN 37831}

\author{D. G. Mandrus}
\affiliation{ Materials Science and Technology Division, Oak Ridge National Laboratory, Oak Ridge, TN 37831}\affiliation{Department of Materials Science and Engineering, University of Tennessee, Knoxville, TN 37996}

\author{B. C. Sales}
\affiliation{ Materials Science and Technology Division, Oak Ridge National Laboratory, Oak Ridge, TN 37831}

\date{\today}
\begin{abstract}
Mo$_3$Sb$_7$ single crystals lightly doped with Cr, Ru, or Te are studied in order to explore the interplay between superconductivity, magnetism, and the cubic-tetragonal structural transition. The structural transition at 53\,K is extremely sensitive to Ru or Te substitution which introduce additional electrons, but robust against Cr substitution. No sign of a structure transition was observed in superconducting Mo$_{2.91}$Ru$_{0.09}$Sb$_7$ and Mo$_3$Sb$_{6.975}$Te$_{0.025}$. In contrast, 3at\% Cr doping only slightly suppresses the structural transition to 48\,K while leaving no trace of superconductivity above 1.8\,K. Analysis of magnetic properties suggests that the interdimer interaction in Mo$_3$Sb$_7$ is near a critical value and essential for the structural transition. All dopants suppress the superconductivity of Mo$_3$Sb$_7$. The tetragonal structure is not necessary for superconductivity.

\end{abstract}

\maketitle

\section{Introduction}

The interplay between structure, magnetism, and superconductivity has been an active topic in the condensed matter community for decades. For a variety of materials, superconductivity is found near a magnetic instability as a function of chemical doping or pressure. This has been well demonstrated in high-Tc cuprates and pniticides. Doping charge carriers, either hole or electron, suppresses the antiferromagnetic order in cuprates or magnetic spin density wave transition in pniticides, and induces superconductivity. The suppression of magnetism and appearance of superconductivity are associated with a structural transition from a high temperature tetragonal phase to a low temperature orthorhombic phase in both cuprates and pniticides. Thus systems with magnetic and/or structural instabilities have been a fertile ground for looking for new superconductors. However, despite a tremendous effort in the last 30 years, the close interplay between magnetism, superconductivity, and structure and the mechanism mediating pairing are still under intense debate.

Despite a low T$_c$ $\sim$ 2.08\,K,\cite{Bukowski2002SSC} the Zintl compound Mo$_3$Sb$_7$ appears to show a close interplay between magnetism, structure, and superconductivity. Cooling across T$_t$\,=\,53\,K, the nearest neighbor (NN) Mo-Mo distance along the crystallographic \emph{c}-axis shortens leading to the structural transition from a high temperature Ir$_3$Ge$_7$-type cubic structure (space group \emph{Im}$\overline{3}$\emph{m}) to a low temperature tetragonal (\emph{I}4\emph{/mmm}) structure.\cite{StructureJP, StructureMRB} This low-temperature tetragonal phase has been suggested to be beneficial for the low-temperature superconductivity.\cite{Intermetallics} It was suggested that the structure transition leads to the formation of spin singlet dimers, valence-bond crystal, and the opening of a spin gap of 120 K. \cite{StructureJP,StructureMRB,Tran2008PRL,ValenceBondXtal,Tran2009JP} The effect of the spin gap on superconductivity might be rather weak, possible spin fluctuations were argued to coexist with superconductivity. \cite{Candolfi2007PRL, Oles2008} A previous study on the transport and magnetic properties under hydrostatic pressure up to 22\,kbar shows that T$_t$ decreases while T$_c$ increases with increasing pressure.\cite{TranPressure} This resembles the pressure dependence of T$_t$ and T$_c$ of V$_3$Si with the A15 structure and \emph{A}Fe$_2$As$_2$ (\emph{A}\,=\,Ca, Sr, and Ba) superconductors.\cite{A15Review,DCJReview,Athena} Moreover, a spin density wave state was suggested under high pressure that competes with superconductivity.\cite{TranPressure}  While high pressure studies on high quality single crystals are needed to confirm the appearance of this spin density wave state, there is ample evidence supporting the close correlation between magnetism, structure, and superconductivity in Mo$_3$Sb$_7$.

The effect of Te/Ru doping at high doping concentrations for Mo$_{3-x}$Ru$_{x}$Sb$_7$ (x$\geq$0.25, 0.50, and 1.0), and Mo$_3$Sb$_{7-x}$Te$_{x}$ (x$\geq$0.30, 1.0, 1.60, and 2.2) has been studied both experimentally and theoretically.\cite{TeHT,RuHT,TeLT,RuLT} Electronic structure calculations suggest a rigid-band behavior, where Te or Ru doping moves the Fermi level toward the valence band edge. This is supported by the observation that Mo$_3$Sb$_7$ changes from a paramagnetic metal to a diamagnetic semiconductor with increasing Te or Ru doping. No superconductivity, structure transition, nor spin gap was observed in heavily doped Mo$_3$Sb$_7$.\cite{TeLT,RuLT} To the best of our knowledge, no detailed study has been performed to explore the correlation between superconductivity, the structure transition and magnetism in lightly doped compositions.

In this paper, we report the effect of light doping on the structure transition, superconductivity, and magnetism of Mo$_3$Sb$_7$. Our results show that the structure transition of Mo$_3$Sb$_7$ is extremely sensitive to charge doping. No sign of a structural transition was observed in the normal state of Mo$_{2.91}$Ru$_{0.09}$Sb$_7$ (3\% Ru-doping) and Mo$_3$Sb$_{6.975}$Te$_{0.025}$ (0.36\% Te-doping), but the structural transition is rather robust against Cr doping. Data analysis discussed below suggests that the interdimer interaction in Mo$_3$Sb$_7$ is close to a critical value and essential for the occurrence of the structural transition. All dopants suppress superconductivity. Our superconducting and cubic Ru- and Te-doped crystals show that the low-temperature tetragonal phase is not a prerequisite for the occurrence of superconductivity.

\section{Experimental Details}

All crystals were grown out of Sb flux as reported before.\cite{YanGrowth} The starting metal powder (Mo, Cr, and Ru) was reduced in flowing Ar balanced with 4\% H$_2$ for 12\,hours at 1000$^o$C before using. To grow doped compositions, Mo or Sb was partially replaced by the desired dopant in the starting materials. The charge/flux ratio of 1:49 was used for all compositions. For Ru and Te-doped compositions, the crystals are well separated from residual flux after decanting at 700$^o$C. However, Cr-doped crystals are normally covered with a layer of Sb flux. This layer of residual flux can be removed mechanically with a surgical blade. Inset of Fig 1 (a) shows a Cr-doped Mo$_3$Sb$_7$ single crystal after cleaning the residual flux. For the growth of Mo$_{3-x}$Cr$_x$Sb$_7$ with x $\geq$ 1 in starting materials, the crystals are small with the largest dimension in submillimeter range and are normally covered with a thick layer of residual flux. These observations suggest that Cr dopant significantly increases the liquidus temperature.

Elemental analysis of the crystals was performed using a Hitachi TM-3000 tabletop electron microscope equipped with a Bruker Quantax 70 energy dispersive x-ray (EDX) system. X-ray diffraction on oriented single crystals and on powder from ground crystals was performed on PANalytical X'Pert Pro MPD powder X-ray diffractometer using Cu K$_{\alpha1}$ radiation. Magnetic properties were measured with a Quantum Design (QD) Magnetic Properties Measurement System in the temperature range 1.8\,K$\leq$T$\leq$\,300\,K. The temperature dependent specific heat and electrical transport data were collected using a 9\,T QD Physical Properties Measurement System.

The density-of-states and additional electrons relative to the stoichiometric compound were calculated from the converged WIEN2K band structure, as performed and presented in Ref. \cite{parker}.  A very large number of k-points - as many as 100,000 in the full Brillouin zone - were used to evaluate the density-of-states, and appropriate k-point convergence tests performed.  The number of additional electrons is determined by integrating the density-of-states, which amounts to the assumption of rigid band behavior in the addition of dopants such as Ru or Te.  For the purposes of determining the experimental doping levels of Ru and Te, each Ru (substituting for Mo) is assumed to donate two additional electrons to the system, as was found by an explicit supercell calculation in Ref. \cite{parker}, while Te (substituting for Sb) was assumed to donate one additional electron to the system, consistent with the observed approach to a semiconducting state induced by Te doping.

\begin{figure} \centering \includegraphics [width = 0.47\textwidth] {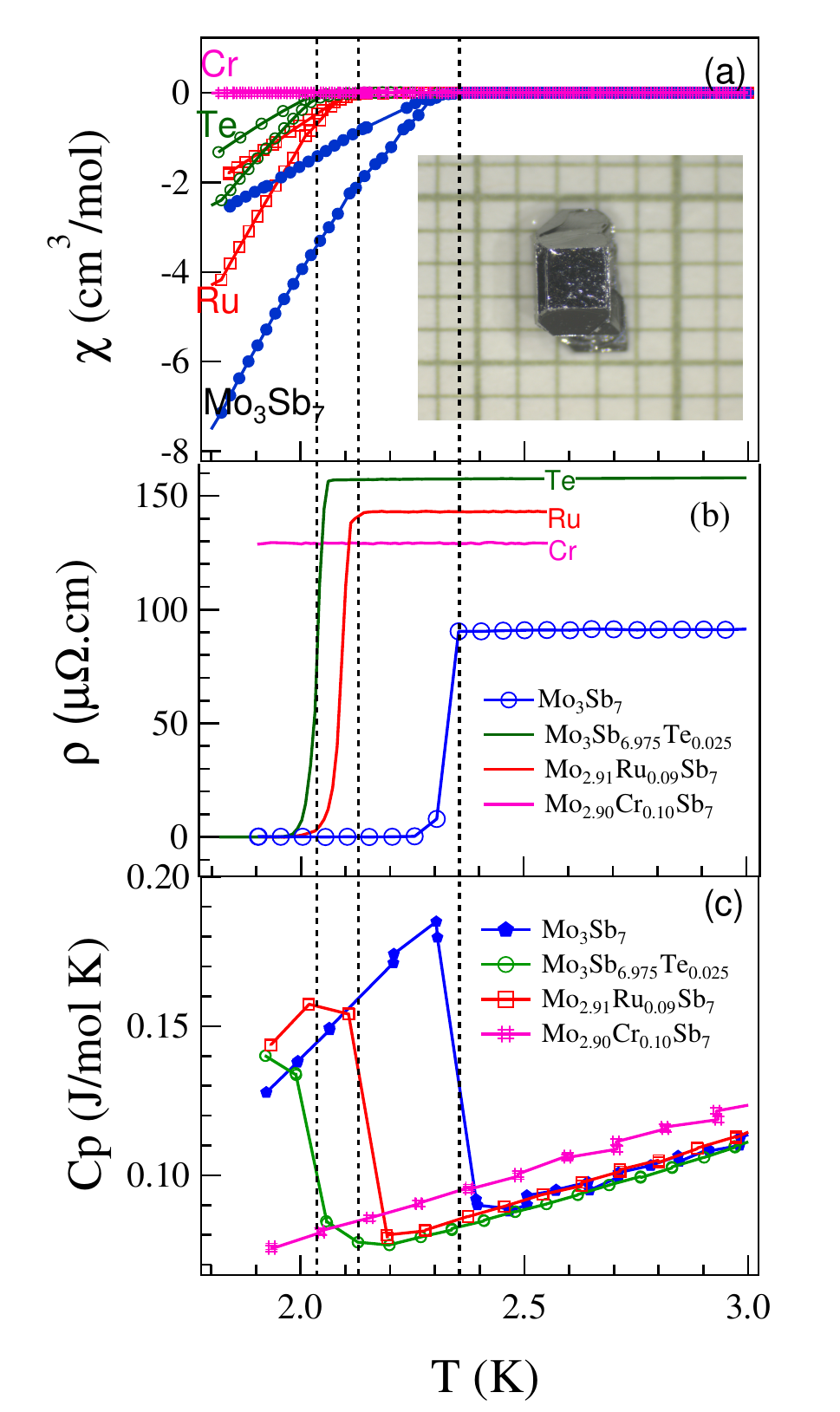}
\caption{(color online) (a) Temperature dependence of magnetic susceptibility around T$_c$ measured under an applied magnetic field of 10\,Oe in both zero-field-cooling and field-cooling modes. Inset shows the photograph of a Mo$_{2.90}$Cr$_{0.10}$Sb$_7$ single crystal on a millimeter grid. (b) Temperature dependence of electrical resistivity around T$_c$. (c) Temperature dependence of specific heat around T$_c$. Data for Mo$_3$Sb$_7$ are reploted from Ref \cite{YanGrowth}. Vertical dashed lines highlight the superconducting transition temperatures. }
\label{Tc-1}
\end{figure}

\begin{figure} \centering \includegraphics [width = 0.47\textwidth] {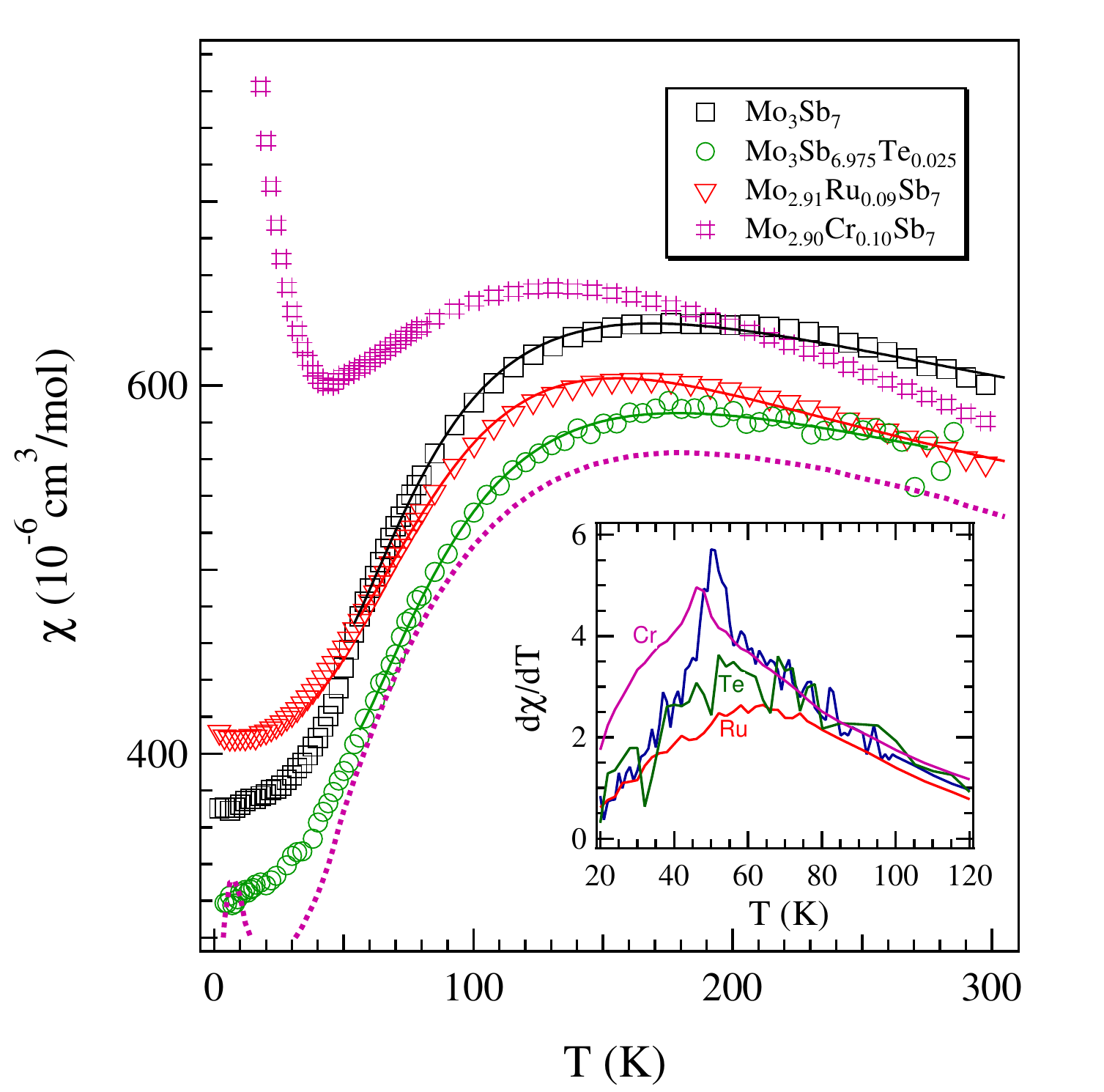}
\caption{(color online) Temperature dependence of magnetic susceptibility measured with a magnetic field of 60 kOe.  The dashed curve shows the magnetic susceptibility of Mo$_{2.90}$Cr$_{0.10}$Sb$_7$ after subtracting the low-temperature Curie-Weiss tail. The solid curves are fittings with the models described in text. The inset shows the temperature derivative of magnetic susceptibility, d$\chi$(T)/dT. d$\chi$(T)/dT for Mo$_{2.90}$Cr$_{0.10}$Sb$_7$ was obtained from the dashed curve. Data for Mo$_3$Sb$_7$ are reploted from Ref\cite{YanGrowth}. }
\label{Mag-1}
\end{figure}

\section{Results and Discussions}

Room temperature x-ray powder diffraction (not shown) confirmed the single phase for all compositions studied. In order to correlate the physical properties with the real dopant content \emph{x} rather than the nominal one \emph{x}$_{nominal}$, elemental analysis was performed with EDX. \emph{x} was obtained by averaging the \emph{x} values measured at 5 different locations on each sample. For Te-doped crystals studied in this work, our EDX doesn't have the resolution to resolve such a small amount of Te. Thus we use the nominal composition in the text. For Ru-doped compound with the nominal composition Mo$_{2.91}$Ru$_{0.09}$Sb$_7$, \emph{x} determined equals to \emph{x}$_{nominal}$. A significant difference between the real and nominal compositions was observed for Cr-doped crystal where Mo$_{2.90}$Cr$_{0.10}$Sb$_7$ was obtained starting with the nominal Mo$_{2.70}$Cr$_{0.30}$Sb$_7$. Considering the observation that Cr-doped crystals cannot be well separated from flux by centrifuging at 700$^o$C, Cr might have a limited solubility in Sb flux at the growth temperatures. If all Cr starting materials are dissolved in Sb flux, the compositional difference suggests a distribution coefficient of Cr \emph{k}$_{eff}$\,=\,C$_L$/C$_S$\,$>$1, where \emph{k}$_{eff}$ is the distribution coefficient, C$_L$ and C$_S$ are the concentration of Cr in the melt and crystals.

Figure\,\ref{Tc-1}(a) shows the temperature dependence of magnetic susceptibility measured at a magnetic field of 10\,Oe in both zero-field-cooling and field-cooling modes. T$_c$ was suppressed from 2.35\,K for Mo$_3$Sb$_7$ to 2.15\,K for Mo$_{2.91}$Ru$_{0.09}$Sb$_7$ and 2.05\,K for Mo$_3$Sb$_{6.975}$Te$_{0.025}$, respectively. No superconductivity was observed for Mo$_{2.90}$Cr$_{0.10}$Sb$_7$ above 1.80\,K. The suppression of superconductivity with doping was further supported by the temperature dependence of the electrical resistivity (Fig.\,\ref{Tc-1}(b)) and specific heat (Fig.\,\ref{Tc-1}(c)).

Figure 2 shows the temperature dependence of magnetic susceptibility, $\chi(T)$, measured under a field of 60\,kOe. The magnetic susceptibility of Mo$_3$Sb$_7$, Mo$_3$Sb$_{6.975}$Te$_{0.025}$, and Mo$_{2.91}$Ru$_{0.09}$Sb$_7$ show similar temperature dependence. Mo$_{2.90}$Cr$_{0.10}$Sb$_7$ exhibits a strong Curie-Weiss (CW) tail below 40\,K  in $\chi(T)$. This is as expected since Cr has a larger localized moment than Mo. The magnetic moment on Cr ions was estimated to be 1.1\,$\mu$ $_B$/Cr by fitting the low-temperature CW tail. It is worth mentioning that the strong CW tail is inherent to the Mo$_{2.90}$Cr$_{0.10}$Sb$_7$ crystals and comes mainly from the isolated Cr spins. The broken Mo-Mo dimers by the Cr substituent may have a small contribution to the low temperature CW tail, as evidenced by the low temperature $\chi(T)$ of Mo$_{2.91}$Ru$_{0.09}$Sb$_7$.

Below 20\,K, $\chi(T)$ curves for other compositions show little temperature dependence which signals the good quality of our crystals. At high temperatures, a broad maximum was observed for all compositions. The broad maximum occurs at approximately T$_{max}$\,=\,175\,K for Mo$_3$Sb$_7$ and Mo$_3$Sb$_{6.975}$Te$_{0.025}$. For the other two compositions with substitution at the Mo site, the broad maximum shifts to lower temperatures.

$\chi(T)$ of Mo$_3$Sb$_7$ shows a rapid drop around T$_t$=53\,K, which leads to a distinct maximum around 53\,K in the d$\chi$ $/$dT (inset of Fig.\,2). For Mo$_{2.90}$Cr$_{0.10}$Sb$_7$, d$\chi$$/$dT was obtained by first subtracting the low temperature CW tail with $\chi$$_{CW}$=C/(T-$\theta$), where C is the Curie constant, and $\theta$ the Weiss constant. A distinct maximum was observed around 48\,K. In contrast, this feature becomes less distinguishable and only a broad maximum was observed around 50\,K in d$\chi$$/$dT of Mo$_3$Sb$_{6.975}$Te$_{0.025}$, and Mo$_{2.91}$Ru$_{0.09}$Sb$_7$.

\begin{figure} \centering \includegraphics [width = 0.47\textwidth] {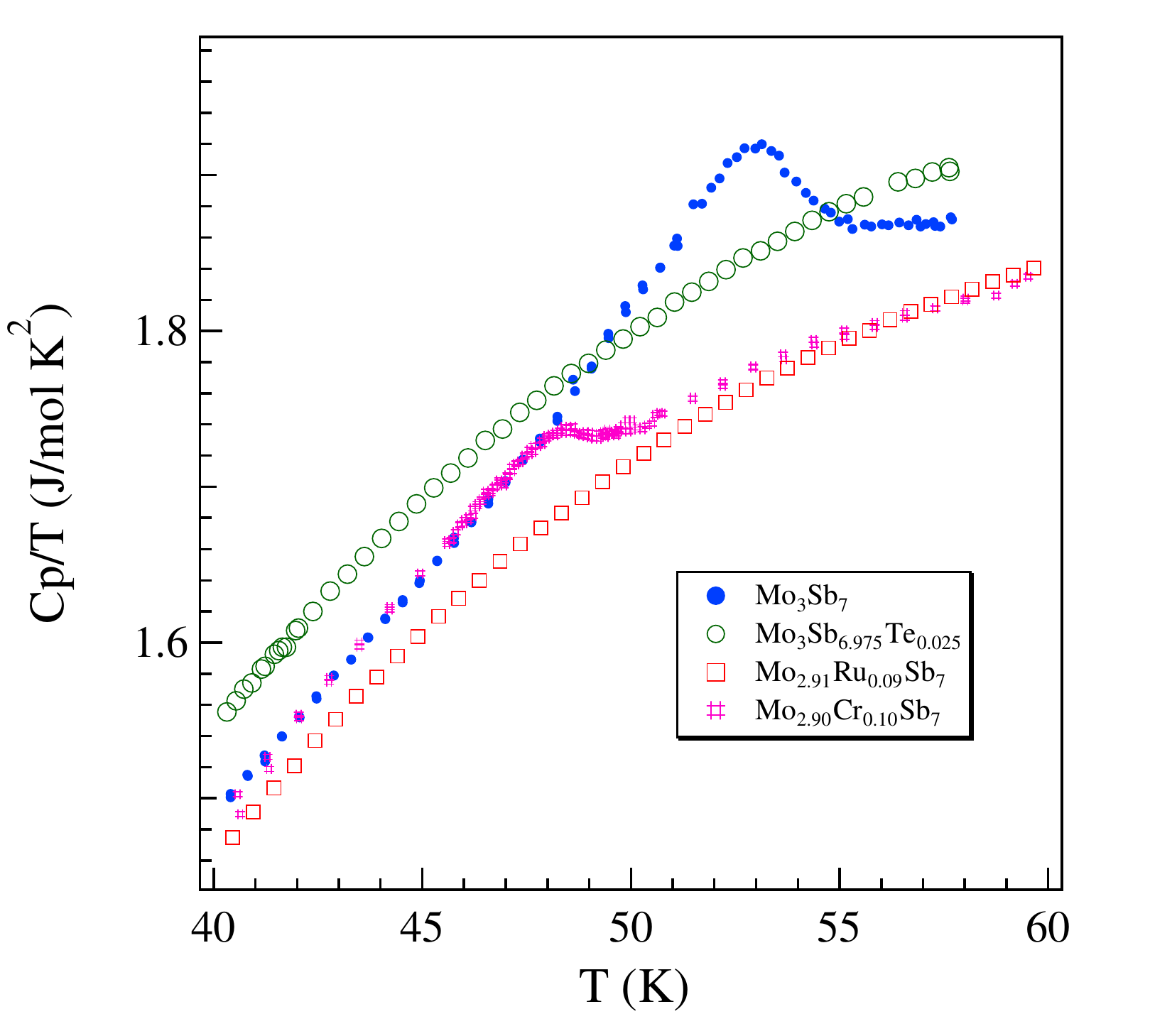}
\caption{(color online) Temperature dependence of specific heat of doped Mo$_3$Sb$_7$ single crystals. Data for Mo$_3$Sb$_7$ are reploted from Ref\cite{YanGrowth}.}
\label{Cp-1}
\end{figure}

Figure \ref{Cp-1} shows the temperature dependence of specific heat plotted as C$_p$(T)/T vs T around the structure transition. The slight difference in the magnitude of specific heat may come from the different Debye temperatures and/or experimental error. A weak lambda anomaly near T*\,=\,53\,K is well resolved for the parent compound. This anomaly becomes smaller in magnitude and shifts to T* $\sim$ 48 K for Mo$_{2.90}$Cr$_{0.10}$Sb$_7$. In contrast, no anomaly was observed in the Cp(T)/T curves in the normal state of  Mo$_3$Sb$_{6.975}$Te$_{0.025}$ and Mo$_{2.91}$Ru$_{0.09}$Sb$_7$. The temperature dependence of Cp(T) and  d$\chi$$/$dT suggests the structural transition disappears in Mo$_3$Sb$_{6.975}$Te$_{0.025}$ and Mo$_{2.91}$Ru$_{0.09}$Sb$_7$.

\begin{figure} \centering \includegraphics [width = 0.47\textwidth] {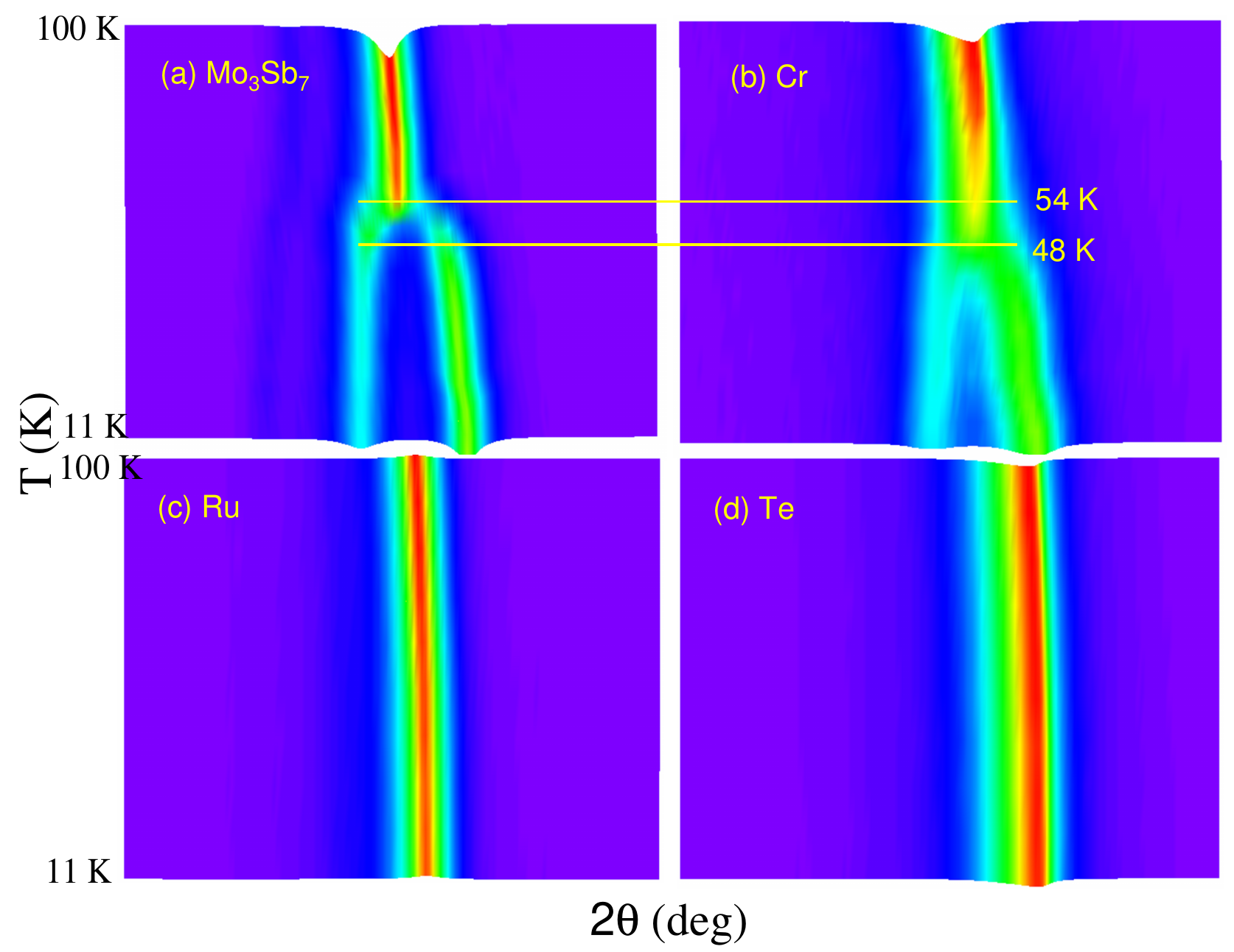}
\caption{(color online) Temperature dependence of (800) reflections for (a) Mo$_3$Sb$_7$,(b) Mo$_{2.90}$Cr$_{0.10}$Sb$_7$, (c) Mo$_{2.91}$Ru$_{0.09}$Sb$_7$, and (d) Mo$_3$Sb$_{6.975}$Te$_{0.025}$. The data were collected on oriented single crystals.}
\label{XRD-1}
\end{figure}

To confirm the doping effect on the structural transition of Mo$_3$Sb$_7$, the evolution with temperature of the cubic 800 peak was studied on oriented single crystals in the temperature interval 10\,K\,$\leq$\,T\,$\leq$ 100\,K. As shown in Fig. \ref{XRD-1}, the (800) peak splits below 54\,K signaling the lowering of the symmetry from cubic to tetragonal for Mo$_3$Sb$_7$. This peak splitting was observed at $\sim$48\,K for Mo$_{2.90}$Cr$_{0.10}$Sb$_7$. In sharp contrast, no peak splitting was observed for Mo$_3$Sb$_{6.975}$Te$_{0.025}$ and Mo$_{2.91}$Ru$_{0.09}$Sb$_7$. The x-ray diffraction experiments provide direct and unambiguous evidence for (1) the absence of the structural transition in Mo$_3$Sb$_{6.975}$Te$_{0.025}$ and Mo$_{2.91}$Ru$_{0.09}$Sb$_7$, and (2) the suppression of structural transition to 48\,K in Mo$_{2.90}$Cr$_{0.10}$Sb$_7$. These results are consistent with other bulk properties. At 11\,K, the a/c ratio was obtained to be 1.002 for both Mo$_3$Sb$_7$ and Mo$_{2.90}$Cr$_{0.10}$Sb$_7$.

The absence of a structural transition in Mo$_{2.91}$Ru$_{0.09}$Sb$_7$  and Mo$_3$Sb$_{6.975}$Te$_{0.025}$ is rather unusual. The shift of T$_{max}$ in $\chi(T)$ curves to lower temperatures signals that the Mo-Mo antiferromagnetic interaction is disturbed by the random distribution of Ru or Cr at Mo site and possible breaking of Mo-Mo dimers. From this point of view, the substitution at Mo site should suppress the structure transition. This is supported by the case of Mo$_{2.90}$Cr$_{0.10}$Sb$_7$ where 3at\% Cr-doping suppresses the structure transition from 53\,K in parent compound to 48\,K. However, no sign of the structure transition was observed above T$_c$ for Mo$_{2.91}$Ru$_{0.09}$Sb$_7$, where similar amount of Mo is replaced by Ru, from the temperature dependence of magnetization, specific heat, electrical resistivity (not shown), and x-ray diffraction. Therefore, disturbing the Mo-sublattice by substitution at Mo-site does not drive the suppression of the structure transition in Mo$_{3}$Sb$_7$. The doping effect of Te on the structural transition further supports the above conclusion.

Substitution of Sb by Te doesn't disturb the Mo-sublattice directly. However, as in Mo$_{2.91}$Ru$_{0.09}$Sb$_7$, no structure transition was observed by X-ray diffraction above 11 K and no anomaly associated with the structure transition was observed in the temperature dependence of magnetization, specific heat and electrical resistivity in the normal state. Thus, Te substitution affects the structure transition in a way similar to that of Ru substitution although they reside at different crystallographic sites. Considering both dopants introduce additional electrons, we suggest that the structural transition is sensitive to electron concentration.

\begin{figure} \centering \includegraphics [width = 0.47\textwidth] {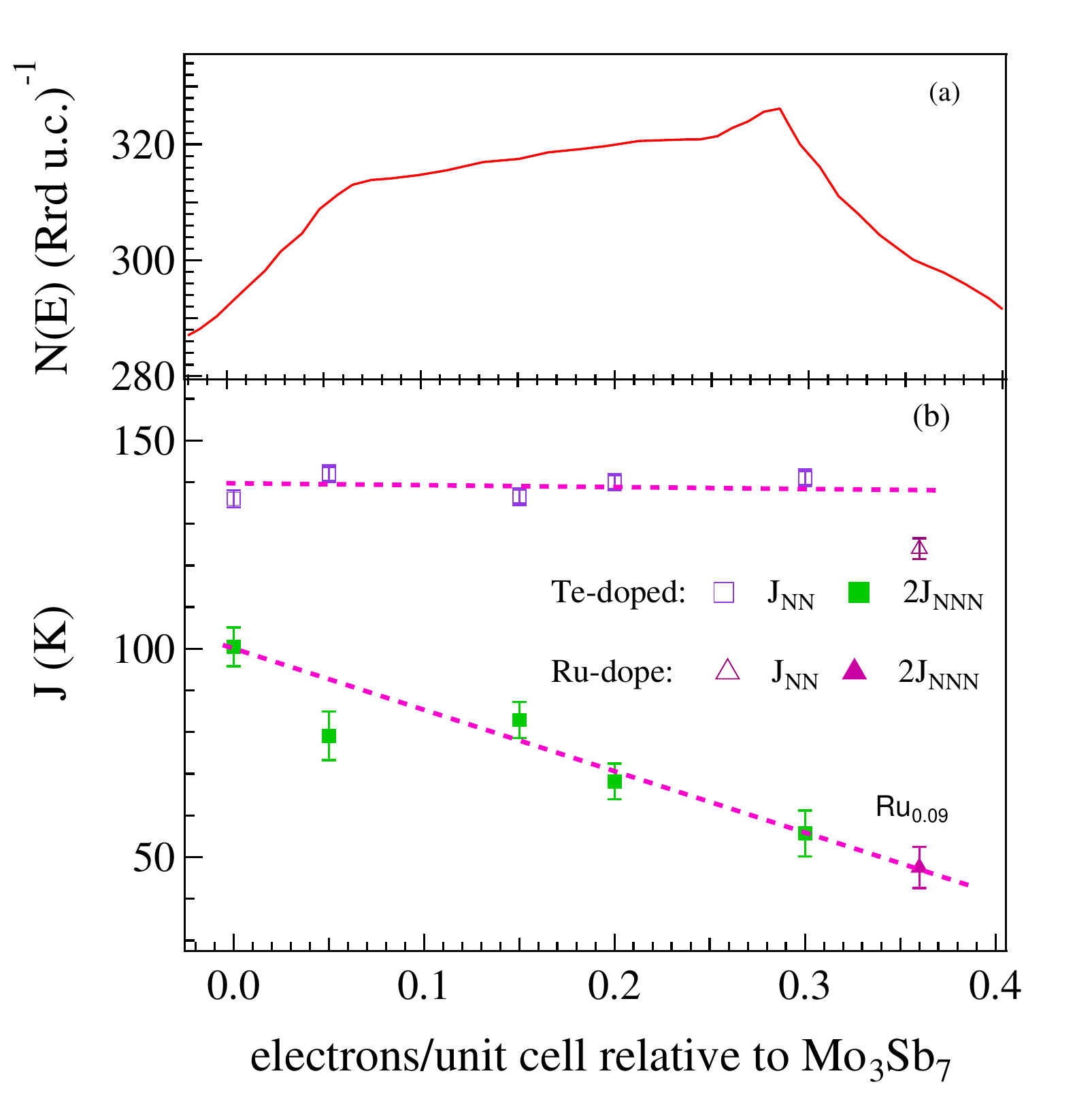}
\caption{(color online) The evolution with extra electrons of (a) the electron DOS at the Fermi level, and (b) J$_{NN}$ (open symbols) and J$_{NNN}$(solid symbols). The number of additional electrons is determined by integrating the density-of-states, which amounts to the assumption of rigid band behavior in the addition of dopants such as Ru or Te.  Dashed lines in (b) are a guide to the eyes.}
\label{DOS-1}
\end{figure}

A band Jahn-Teller effect was proposed to induce the structural instability in cubic superconductors such as A15 compounds.\cite{A15Review} However, our band structure calculations for Te and Ru doped compositions suggest that light doping induces little change of the degeneracy and curvature of the valence bands, consistent with previous reports.\cite{RuLT, TeLT} Thus a band Jahn-Teller effect is unlikely to drive the structural transition in Mo$_3$Sb$_7$. The density-of-states with doping is presented in Fig. \ref{DOS-1}, and was calculated from the converged WIEN2K band structure, as performed and presented in Ref. \onlinecite{parker}. The electron DOS at the Fermi level increases slightly with Te or Ru substitution in the doping range studied in this work. No dramatic change of the electron DOS at the Fermi level is induced by the light doping.

To further understand how Ru or Te dopants affect the structural transition, the magnetic susceptibility data are analyzed in more detail as a previous study measuring $^{121/123}$Sb nuclear quadrupole resonance and muon spin relaxation of Mo$_3$Sb$_7$ suggested the importance of the interdimer magnetic interactions.\cite{ValenceBondXtal} The Mo-Mo magnetic interactions between the nearest neighbors (\emph{J}$_{NN}$) (i.e., the intradimer interaction) and next nearest neighbors (\emph{J}$_{NNN}$) were estimated by fitting the $\chi$(T) data above T$_t$ with the mean-field modification of the Bleaney-Bowers equation:\cite{Tran2008PRL, Bowers, Ba3Cr2O8, JMC}

\begin{equation}
\chi(T)=\chi_0+\frac{N_A{\mu_B }^{2}g^2}{k_BT(3+exp(2J/k_BT)+J'/k_BT)}
\end{equation}

\noindent where $\chi_0$ is a temperature independent term, N$_A$ is the Avogadro number, $\mu_B$ is the Bohr magneton, k$_B$ is the Boltzmann constant, \emph{J}\,=\,\emph{J}$_{NN}$ is the intradimer interaction, and \emph{J'} is the interdimer interaction beyond the nearest Mo-Mo neighbors. The interdimer interaction includes the interaction, J$_{NNN}$, between next nearest neighbours  with a Mo-Mo distance of 4.642 ${\AA}$ and the interaction J$_1$ between the chains with a Mo-Mo distance of 5.220 ${\AA}$.\cite{YanGrowth}  Thus the interdimer interaction can be written as J'=8J$_{NNN}$+4J$_1$. With the assumption of J is proportional to the Mo-Mo distance, J$_1$ can be estimated as 0.88J$_{NNN}$. \cite{Tran2008PRL} The obtained magnetic interactions are shown in Fig.\,\ref{DOS-1} as a function of extra electrons. The following features are noteworthy:(1) the intradimer interaction \emph{J}$_{NN}$ is suppressed by partial substitution of Mo with Ru, but not by Sb with Te. This agrees with an intuitive picture that Te substitution disturbs only the Sb sublattice. \emph{J}$_{NN}$ shows little dependence on Te doping, indicating the absence of structural transition in doped compositions is not driven by \emph{J}$_{NN}$; (2) substitution at either Mo or Sb site rapidly suppresses \emph{J}$_{NNN}$; and (3) a larger suppression of \emph{J}$_{NNN}$ ($\sim$50\%) is in contrast to a smaller suppression of \emph{J}$_{NN}$ ($\sim$10\%) in Mo$_{2.91}$Ru$_{0.09}$Sb$_7$. Considering that the structural transition disappears in both Te and Ru doped compositions, we suggest that the structural instability is very sensitive to \emph{J}$_{NNN}$. The absence of the structural transition in 0.36\% Te-doped composition signals that \emph{J}$_{NNN}$ in Mo$_3$Sb$_7$ is near a critical value and any electron doping which suppresses \emph{J}$_{NNN}$ would remove the structural instability.

For the $\chi$(T) of Mo$_{2.90}$Cr$_{0.10}$Sb$_7$, we always observe the anomaly around 48\,K in d$\chi$/dT curve when we fit the low-temperature CW term with different parameters. However, \emph{J}$_{NN}$ and \emph{J}$_{NNN}$ determined using Equ. (1) depend on the low-temperature CW term. Therefore, unfortunately, we cannot extract reliable J from fitting $\chi$(T) of Mo$_{2.90}$Cr$_{0.10}$Sb$_7$.

The temperature dependence of magnetization, electrical resistivity and specific heat shown in Fig.\,1 clearly demonstrates that all dopants suppress T$_c$. The absence of superconductivity in Mo$_{2.90}$Cr$_{0.10}$Sb$_7$ above T$\geq$1.8\,K suggests that the superconductivity in Mo$_3$Sb$_7$ is sensitive to magnetic dopants.\cite{Maple} Together with the fact that light Ru and Te-doping suppresses superconductivity but increases the DOS at E$_F$, we can conclude that the superconductivity in Mo$_3$Sb$_7$ is sensitive to disorder and/or magnetic dopants as in cuprates. This makes it difficult to understand how the superconductivity reacts to the suppressed \emph{J}$_{NNN}$ by Te or Ru doping. The occurrence of superconductivity in the cubic phase of Mo$_3$Sb$_{6.975}$Te$_{0.025}$ and Mo$_{2.91}$Ru$_{0.09}$Sb$_7$ and the tetragonal phase of Mo$_3$Sb$_7$, and the absence of superconductivity in the tetragonal phase of Mo$_{2.90}$Cr$_{0.10}$Sb$_7$ observed in this study suggest that the tetragonal phase is not necessary for the superconductivity.

\section{Summary}

In summary, the effects of light doping of Mo$_3$Sb$_7$ with Ru, Cr, or Te have been studied in single crystals grown by a self-flux technique. All dopants suppress superconductivity. Superconductivity in Mo$_3$Sb$_7$ is sensitive to magnetic dopants and disorder but can exist in both cubic and tetragonal phases. The structural transition is extremely sensitive to additional electrons introduced by Te or Ru substitution. Analysis of the magnetic susceptibility data suggests that the interdimer magnetic interaction, \emph{J}$_{NNN}$, is essential for the structural transition and is close to a critical value in Mo$_3$Sb$_7$. Our study highlights the importance of magnetism in the structure transition of Mo$_3$Sb$_7$. The sensitivity of the structural transition to electron doping and superconductivity to substitution suggests that hydrostatic pressure would be a cleaner tool to tune the ground state and to explore the correlation between structure, magnetism, and superconductivity in Mo$_3$Sb$_7$.

\section{Acknowledgments}
This research was supported by the U.S. Department of Energy, Office of Science, Basic Energy Sciences, Materials Sciences and Engineering Division.

\end{document}